\documentclass[twocolumn]{aastex631}

\usepackage{esdiff}
\usepackage{lmodern}
\usepackage{latexsym}
\usepackage{amsmath}
\usepackage{amssymb}
\usepackage{amsbsy}
\usepackage{amsthm}
\usepackage{amsfonts}
\usepackage{mathrsfs}
\usepackage{bm}
\usepackage{sansmath}
\usepackage{relsize}
\usepackage{caption2}
\usepackage{graphicx}
\usepackage[utf8]{inputenc} 
\usepackage[T1]{fontenc}
\usepackage{epstopdf}

\shorttitle{Hot Accretion Flow around Bondi Radius}
\shortauthors{Mosallanezhad et al.}
\graphicspath{{./}{figures/}}

\begin{document}

\title{Numerical Simulation of Hot Accretion Flow around Bondi Radius}

\correspondingauthor{Amin Mosallanezhad, De-Fu Bu}
\email{mosallanezhad@xjtu.edu.cn, dfbu@shao.ac.cn}

\author[0000-0002-4601-7073]{Amin Mosallanezhad}
\affiliation{School of Mathematics and Statistics, Xi'an Jiaotong University, Xi'an, Shaanxi 710049, PR China}

\author[0000-0002-0427-520X]{De-Fu Bu}
\affiliation{Shanghai Astronomical Observatory, Chinese Academy of Sciences, Shanghai 200030, China}

\author[0000-0002-3434-3621]{Miljenko \v{C}emelji\'{c}}
\affiliation{Nicolaus Copernicus Astronomical Center, Polish Academy of Sciences, Bartycka 18, 00-716 Warsaw, Poland}
\affiliation{Academia Sinica, Institute of Astronomy and Astrophysics, P.O. Box 23-141, Taipei 106, Taiwan}
\affiliation{Research Centre for Computational Physics and Data Processing, Institute of Physics, Silesian University in Opava, Bezru\v{c}ovo n\'am.~13, CZ-746\,01 Opava, Czech Republic}

\author[0000-0003-3345-727X]{Fatemeh Zahra Zeraatgari}
\affiliation{School of Mathematics and Statistics, Xi'an Jiaotong University, Xi'an, Shaanxi 710049, PR China}

\author[0000-0002-0427-520X]{Yang Hai}
\affiliation{Shanghai Astronomical Observatory, Chinese Academy of Sciences, Shanghai 200030, China}

\author[0000-0003-3468-8803]{Liquan Mei}
\affiliation{School of Mathematics and Statistics, Xi'an Jiaotong University, Xi'an, Shaanxi 710049, PR China}



\begin{abstract}

Previous numerical simulations have shown that strong winds can be produced in the hot accretion 
flows around black holes. Most of those studies focus only on the region close to the central black
 hole, therefore it is unclear whether the wind production stops at large radii around Bondi radius.
Bu et al. 2016 studied the hot accretion flow around the Bondi radius in the presence of nuclear 
star gravity. They find that when the nuclear stars gravity is important/comparable to the black hole 
gravity, winds can not be produced around the Bondi radius. However, for some galaxies, the nuclear 
stars gravity around Bondi radius may not be strong. In this case, whether winds can be produced 
around Bondi radius is not clear. We study the hot accretion flow around Bondi radius with and 
without thermal conduction by performing hydrodynamical simulations. We use the virtual particles 
trajectory method to study whether winds exist based on the simulation data. 
Our numerical results show that in the absence of nuclear stars gravity, winds can 
be produced around Bondi radius, which causes the mass inflow rate decreasing inwards. 
We confirm the results of Yuan et al. which indicates this is due to the mass loss of gas 
via wind rather convectional motions.

\end{abstract}

\keywords{accretion, accretion discs --- hydrodynamics --- methods: numerical --- galaxies: active -- galaxies: nuclei}


\section{Introduction} \label{sec:intro}

Hot accretion flow with low mass accretion rate is an important and distinguished 
class of accretion disks. Compared to the well-known standard thin disk (cold accretion mode),
hot accretion flow has a lower density, as well as higher temperature 
and scale height. Consequently, the radiative efficiency drops dramatically
so this model is called radiatively inefficient accretion flow (RIAF). 
Hot accretion flow model is arguably the standard model of low-luminosity 
active galactic nuclei (LLAGNs) in the majority of galaxies in the nearby universe (see 
e.g. \citealt{Ho 2008}; \citealt{Antonucci 2012}), and the hard/quiescent states of black hole 
X-ray binaries (BHXBs) as well (see, e.g. \citealt{Narayan and McClintock 2008}; 
\citealt{Belloni 2010}; \citealt{Yuan and Narayan 2014}).  

Discovery of strong wind is one of the most important advantages in our understanding of hot accretion 
flows in recent years (see, e.g., \citealt{Stone et al. 1999}; \citealt{Stone and Pringle 2001}; 
\citealt{Yuan et al. 2012a, Yuan et al. 2012b}; \citealt{Narayan et al. 2012}; \citealt{Li et al. 2013}; 
\citealt{Gu 2015}; \citealt{Yuan et al. 2015}). The existence of wind in hot accretion flow is also 
confirmed by the 3 Ms Chandra observation of the supermassive black hole (SMBH) in our 
galactic centre, Sgr A* (\citealt{Wang et al. 2013}). Wind not only plays a significant role
in AGN feedback (e.g., \citealt{Ostriker et al. 2010}) but it also brings us important insights and 
questions regarding the physics of the black hole accretion.

\citealt{Yuan et al. 2015} had extensively studied the properties of wind including terminal velocity 
and mass flux of the wind for the hot accretion flow at small scale, $ r < 10^{3} r_{\rm s} $, 
where $ r_{\rm s} $ is the Schwarzschild radius. They used virtual trajectory test particle method. 
In principle, a trajectory is obtained by connecting the positions of the same test particle at 
different times. This concept is related to the Lagrangian description of fluid and is different from 
the streamline. Based upon this method they showed that the mass accretion rate decreases 
inward due to mass loss of wind rather than convection. According to their results, the poloidal velocity 
of wind roughly follows $ v_{\rm p, wind} \approx (0.2-0.4) v_{\rm _K}(r) $, where $ v_{\rm _K}(r) $ 
represents the Keplerian velocity at radius $ r $.  In addition, they found that the mass flux 
of wind can be described as
\begin{equation} \label{mdot_wind}
	\dot{M}_{\rm wind} \approx \dot{M}_{\rm BH} (r / 20 r_{\rm s}), 
\end{equation}
where $ \dot{M}_{\rm BH} $ is the mass accretion rate at the black hole horizon. 
The above equation mainly means that the wind comes from large radii.
Therefore, the important question will be: how far the wind can be produced?~or equivalently,
where is the upper limit of the radius that the equation (\ref{mdot_wind}) can be applied to?

To answer such questions, we first need to introduce some physical radii related to the 
accretion process. For the spherically symmetric flow, in the absence of the rotation and 
radiation, the Bondi radius (\citealt{Bondi 1952}) will be defined as, 
\begin{equation} \label{Bondi_radius}
	r_{\rm _B} = \frac{GM}{c_{\rm s,\infty}^{2}},
\end{equation}
where $ M $ is the black hole mass, $ G $ is the gravitational constant, and $ c_{\rm s,\infty} $ 
represents the sound speed of the gas at infinity. From this radius towards 
the black hole the negative gravitational energy dominates over the thermal energy of
the gas. However, in real accretion flow, the gas has some amount of angular 
momentum and rotates around the central black hole. Therefore, we can define a characteristic 
radius at which the centrifugal and gravitational forces balance each other. The centrifugal 
radius is given as,
\begin{equation} \label{centrifugal_radius}
	r_{\rm _C} = \frac{\ell^{2}}{GM},
\end{equation}  
where $ \ell $ is the angular momentum per unit mass. This radius is 
larger than Schwarzschild radius and can be less than Bondi one, i.e., 
$ r_{\rm s} \ll r_{\rm _C} < r_{\rm _B} $. 
Existence of any mechanism for driving angular momentum tends to give rise 
the inflowing gas form a rotating accretion disc before falling down 
on to the black hole. It is well known that in a real accretion flow the angular 
momentum is transferred by Maxwell stress associated with Magnetohydrodynamic 
(MHD) turbulence driven by Magneto-rotational instability (MRI; \citealt{Balbus and Hawley 1998}). 
In hydrodynamical (HD) simulations, it is common to mimic the effect of the 
magnetic stress by adding viscosity terms for both driving angular momentum 
outwards as well as producing heat. 

A good model for studying the dynamics and the structure of the accretion gas onto a SMBH 
should cover a wide range of spatial scales, from the inner region, where an accretion disc forms, 
to the the region outside the Bondi radius, where the accretion process originates in the first steps. So far,
several HD and MHD simulations have been done to connect large and small scales (see, i.e.,
\citealt{Proga and Begelman 2003a, Proga and Begelman 2003b}; \citealt{Li et al. 2013}; 
\citealt{Bu et al. 2016a, Bu et al. 2016b}; \citealt{Inayoshi et al. 2018, Inayoshi et al. 2019}).
For instance, \citealt{Bu et al. 2016a, Bu et al. 2016b} studied the accretion flow around Bondi radius. 
They found that by including the gravity of nuclear stars, the physics can be totally changed. 
More precisely, they showed that in the presence of nuclear star gravity, the winds can not be 
produced locally. They also did some tests and found that when the nuclear star gravity is 
excluded, the winds can be generated locally around Bondi radius.
On the other hand, \citealt{Inayoshi et al. 2018} performed two-dimensional 
HD simulations of hot accretion flow at a range about $ 10^{-2} r_{\rm _B} \leq r \leq 50 r_{\rm _B} $. 
They found a global steady accretion solution with two distinguished regimes: (1) the outer rotational equilibrium 
region around Bondi radius follows the density profile of $ \rho \propto (1 + r_{\rm _B} / r)^{3/2} $ 
with subsonic gas motion; (2) the inner solution where the geometrically thick torus follows 
$ \rho \propto r^{-1/2} $. Based upon the density and the mass accretion rate ($ \dot{M} \propto r $) 
profiles of inner part, they argued that the physical properties of this region are consistent with the 
convection-dominated accretion flow (CDAF) model proposed by \citealt{Narayan et al. 2000}.
More precisely, since they did not find wind in their solutions, they claimed that the adiabatic 
inflow-outflow solution (ADIOS; \citealt{Blandford and Begelman 1999, Blandford and Begelman 2004}; 
\citealt{Begelman 2012}) is not the main reason for decreasing mass accretion rate at such large radii. 
Therefore, they concluded that the convection causes the mass accretion rate decreases inward.  

The motivations for performing the HD simulations of this paper are threefold. 
First, we want to check whether or not the inward decrease of mass accretion rate 
is due to the convection. The second purpose is to study the hot accretion 
flow at large radii to investigate how far the wind can move outward. 
The third motivation is the study of the effects of the thermal conduction on the wind. This is mainly because 
for systems with extremely low accretion rate, such as our galactic center Sgr A* and M87 galaxy, 
the accretion flows are weakly collisional. The electron collisional mean free path then
can be much larger than its Larmor radius. Consequently, the conduction can significantly 
influence the dynamics of the accretion flow and transport energy from the inner to 
the outer regions (\citealt{Johnson and Quataert 2007}; \citealt{Quataert 2008}).
In this paper, we revisit \citealt{Inayoshi et al. 2018} by performing numerical HD simulations
with some modifications to extensively study the detailed properties of 
wind at large radii. Following \citealt{Yuan et al. 2015}, we use the much more precise 
trajectory analysis of virtual test particles based on our simulation data to study whether does wind exist
(see section \ref{sec:trajectory_method} for more details). 

The main structure of the paper is as follows. In Section \ref{sec:method}, we will describe 
the basic HD equations, simulation method, and initial and boundary conditions. The results will be 
presented in Section \ref{sec:results} and we briefly overview of the trajectory method we use to 
analyze the simulation data (subsection \ref{sec:trajectory_method}). We then summarize 
our work in Section \ref{sec:summary}.

\section{Method} \label{sec:method}

We perform axisymmetric two-dimensional (2D) HD simulations
using the publicly available numerical simulation package {\tt PLUTO} \footnote{http://plutocode.ph.unito.it}, 
which is a finite-volume/finite-difference, shock-capturing code designed to integrate a 
system of conservation lows based on conservative Godunov scheme (\citealt{Mignone et al. 2007}). 
Our basic simulation setup builds upon \citealt{Inayoshi et al. 2018, Inayoshi et al. 2019}. 
In the following subsections, we outline our numerical model and the differences from 
those aforementioned simulations.

\subsection{Basic equations}

To compute the structure and the evolution of an accretion flow with low mass accretion rate 
around SMBH at large radii, we solve the basic HD equations, including the equation of continuity,

\begin{equation} \label{eq:continuity}
	\diff{\rho}{t} + \rho \nabla \cdot \bm{v} = 0,
\end{equation}
the momentum balance equation,
\begin{equation} \label{eq:momentum}
	\rho \diff{\bm{v}}{t} = - \nabla p - \rho \nabla \psi + \nabla \cdot \bm{\sigma},
\end{equation}
and the energy equation,

\begin{equation} \label{eq:energy}
	\rho \diff{e}{t} = - p \nabla \cdot \bm{v} + (\bm{\sigma} \cdot \nabla) \bm{v} - \nabla \cdot \bm{Q}.
\end{equation}
\\
In the above equations, $ \rho $, $ \bm{v} $, $ p $, $ \psi $, and $ e $ are the density, 
velocity, gas pressure, gravitational potential, and internal energy per unit mass, respectively. 
Here, we only take into account the gravity of the central black hole and 
neglect the nuclear stars gravity\footnote{Based on Jaffe model the amount of stellar mass contained 
within a sphere of  30 Bondi radius is about $ 8.6 M_{\rm BH} $ for a galaxy of 
$ 10^{11} M_{\odot} $ with a black hole mass of $ M_{\rm BH} = 10^{8} M_{\odot} $.}. 
Therefore, the black hole potential can be adopted as $ \psi = - GM / r $. 
$ \bm{\sigma} $ is the viscous 
stress tensor and $ \bm{Q} $ in the last term of the right-hand side of energy equation 
represents the thermal conduction. The radiative losses and AGN feedback 
are not considered in this study. The Lagrangian/comoving derivative presented in 
the first terms of the above set of equations is given by 
$ \mathrm{d} / \mathrm{d} t  \equiv \partial / \partial t + \bm{v} \cdot \nabla $. 
We assume the equation of state of ideal gas in the form of $ p = \left( \gamma - 1 \right) \rho e $ 
and set $ \gamma = 5/3 $.

In real accretion flow, the angular momentum is transferred by the Maxwell
stress associated with MHD turbulence driven by MRI (\citealt{Balbus and Hawley 1998}). 
Although we do not include magnetic field in our HD simulations, we assume the viscous 
stress tensor, $ \bm{\sigma} $, to mimic its effects for driving angular momentum and producing 
dissipation heat (see e.g., \citealt{Yuan et al. 2012a}; \citealt{Bu et al. 2016a}). 
The components of the viscous stress tensor are given by,
\begin{equation} \label{sigma_ij}
	\sigma_{ij} = \rho \nu \left[  \left( \frac{\partial v_{j}}{\partial x_{i}} + \frac{\partial v_{i}}{\partial x_{j}} \right) 
	- \frac{2}{3} \left( \nabla \cdot \bm{v} \right) \delta_{ij} \right], 
\end{equation} 
where $ \nu $ is the kinematic viscosity coefficient and $ \delta_{ij} $ is the usual Kronecker delta
\footnote{Note that the bulk viscosity is neglected in our definition for viscous stress tensor.}. 
We adopt the standard $ \alpha $-prescription of viscosity (\citealt{Shakura and Sunyaev 1973}) as,

\begin{equation} \label{nu_vis}
	\nu = \alpha \frac{c_{\rm s}^{2}}{\Omega_\mathrm{K}},
\end{equation} 
 where $ \alpha $ is the viscosity parameter, $ c_{s} $ is the sound speed, and 
$ \Omega_\mathrm{K} ( \equiv \sqrt{GM/r^{3}} ) $ is the Keplerian angular velocity.
Following \citealt{Inayoshi et al. 2018}, the viscosity parameter is defined as,
\begin{equation} \label{alpha_vis}
	\alpha = \alpha_{_0}  \left\{ \exp \left[ - \left( \frac{\rho_{\rm crit}}{\rho} \right)^{2} \right] 
	+ {\rm max} \left( 0, -\frac{\partial \ln \ell}{\partial \ln r} \right) \right\},
\end{equation} 
where $ \alpha_{_0} $ is the strength of the viscosity and $ \rho_{\rm crit} $ 
represents the threshold of the density above which the viscosity will be 
turned on (see equation (\ref{rho_crit}) for more details). The second term 
in equation (\ref{alpha_vis}) is considered to 
achieve steady state accretion flow, since our 2D simulations cannot 
capture the three-dimensional (3D) effect of rotational instability. 
This form of the viscosity causes the viscous process becoming active in the 
disk region where the angular velocity has a significant fraction of the Keplerian, 
i.e., $ r \lesssim r_{\rm _C} $. 

Thermal conduction is also expected to modify the accretion flow structure 
significantly. Here, we explain the definition of thermal conduction 
which is present in models B and C (see table \ref{tab:models}). As we mentioned in
the introduction part, one of the purpose of this paper is to carry out simulations to study the effects 
of thermal conduction on wind at large radii. Motivated by the results of \citealt{Sharma et al. 2008}, 
for our runs with conduction, the thermal conduction term $ - \nabla \cdot \bm{Q} $ is 
added to the right-hand side of the energy equation. In purely HD limit, the heat flux 
can be written as $ \bm{Q} = - \kappa \nabla T $, where $ \kappa $ is the thermal 
diffusivity. In a one-temperature structure, as in our case, temperature can be evaluated 
as $ T = \mu m_{\rm p} p / (k_{\rm _B} \rho) $ where $ k_{\rm _B} $ is the Boltzmann constant, 
$ m_{\rm p} $ is the proton mass, and $ \mu $ is the mean molecular weight.
Following \citealt{Sharma et al. 2008} and \citealt{Bu et al. 2016} for the form of thermal diffusivity, we will adopt
\begin{equation}
	\chi =  \frac{\kappa T}{p} = \alpha_{\rm c} \sqrt{GM r},
\end{equation}
where $ \alpha_{\rm c} = [0.2-2] $ is the dimensionless conductivity.
To prevent the conduction time step from being excessively short, we reduce the conductivity 
in the ambient medium. To do so, we limit the conductivity so that the time scale of conduction, 
$ t_{\rm cond} = r^{2} / \chi $, would be longer than $ 0.1 $ times the dynamical time-scale, $ t_{\rm dyn} = (r^{3} / GM)^{1/2} $. 
This also can effectively mimic the effect of saturated conduction in extremely hot plasma regions.

\subsection{Numerical method} \label{sec:numerical_method}

\begin{figure}[ht!]
\includegraphics[width=0.48\textwidth]{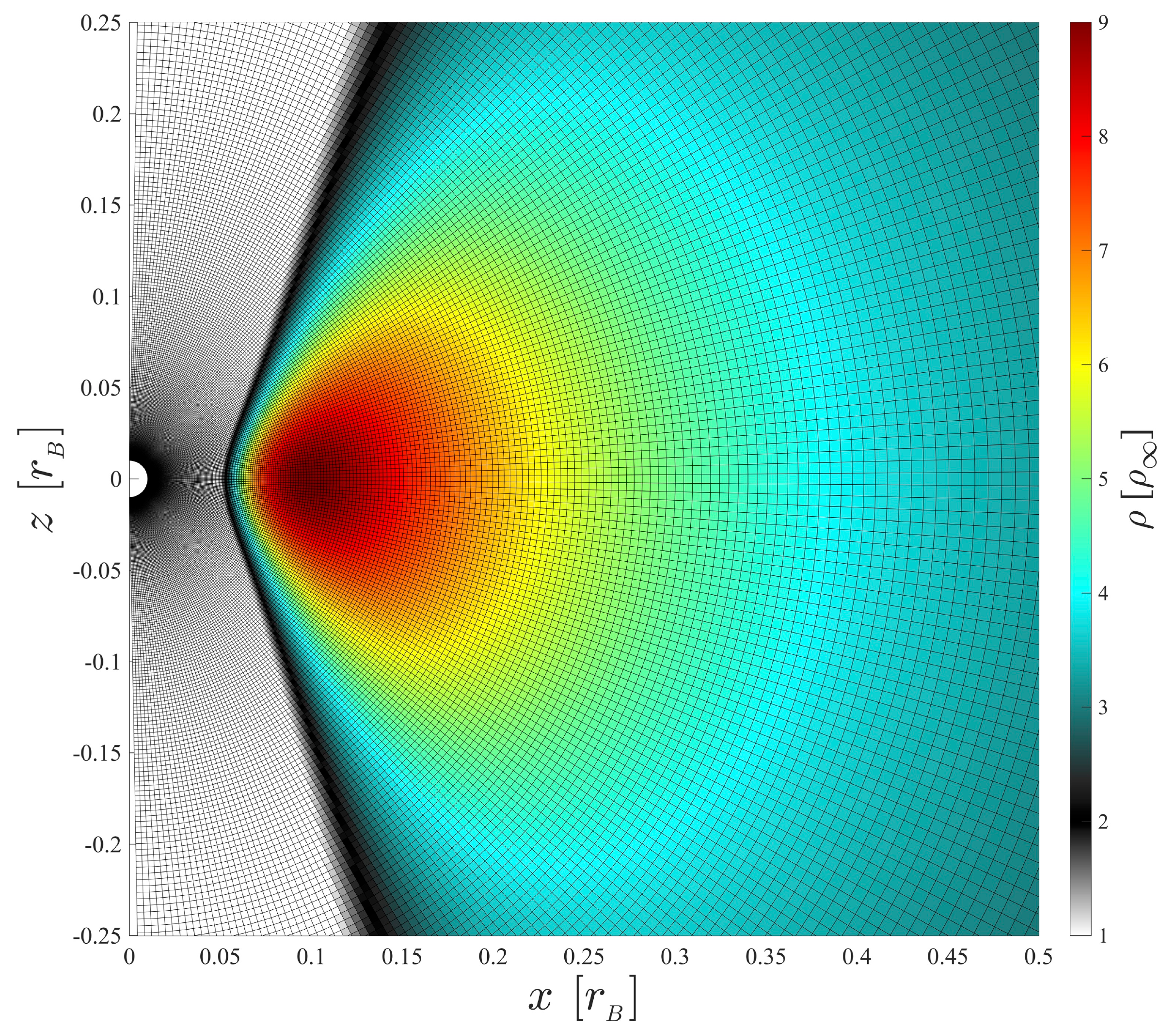}
\centering
\caption{The two-dimensional density distribution of the initial equilibrium torus overlaided with the grid structure 
of our simulations. The maximum density is located at $ 0.1\, r_{_{\rm B}} $. In $ \theta $-direction, 
we set $ N_{\theta} = 200 $ equally spaced grids while in the $ r $-direction, logarithmic grid has 
be adopted which has the advantage of preserving the cell aspect ratio at any distance from the 
origin. The values of $ N_{r} $ and $ r_{\rm max} $ are evaluating so that we have squared cells 
in each direction. \label{contour}}
\end{figure}

To solve the system of equations (\ref{eq:continuity})-(\ref{eq:energy}), we adopt
spherical polar coordinates ($ r, \theta, \phi $). Simulation settings are almost the 
same as those in \citealt{Inayoshi et al. 2018, Inayoshi et al. 2019} except 
some modifications explained in this subsection. For all the runs presented here, 
we assume the black hole mass as $ M = 10^{8} M_{\odot} $, where $ M_{\odot} $ is
the solar mass. The 2D computational domain is set to be $ r_{\rm min} \le r \le r_{\rm max} $ 
and $ \varepsilon \le \theta \le \pi - \varepsilon $, 
where $ \varepsilon $ is considered to be a very small value, to avoid the numerical 
singularity near the polar axis ($ \varepsilon = 10^{-2} $). Unlike \citealt{Inayoshi et al. 2018}, 
we fix the inner radial domain at $ r_{\rm min} = 10^{-2}\, r_{_{\rm B}} $ 
and divide the $ r - \theta $ plane into zones as follows: (1) in the $ \theta $-direction, 
we set $ N_{\theta} = 200 $ equally spaced grid cells; (2) in $ r $-direction, logarithmic grid 
with $ N_{r} = 512 $ zones is adopted. The radial logarithmic grid has the 
advantage of preserving the cell aspect ratio at any distance from the origin. 
The grid cells are plotted in Figure \ref{contour}.
The endpoint in the radial direction, $ r_{\rm max} $, with approximately squared 
cells is determined as,

\begin{equation}
	\log_{_{10}} \left( \frac{r_{\rm max}}{r_{\rm min}} \right) = N_{r}  
	\log_{_{10}} \left( \frac{2 + \Delta \theta}{2 - \Delta \theta} \right).
\end{equation} 

The above formula gives us $ r_{\rm max}  = 30\, r_{_{\rm B}} $. In addition, preserving 
the cell aspect ratio at any distance is necessary for correctly calculating flux at the left 
and right faces of each individual cell used in finite volume method. Using radial logarithmic 
grid can also guarantee a good resolution near the inner region of our computational 
domain where the viscosity becomes important. For the reconstruction of the characteristic 
variables within a cell, we adopted the third-order piecewise parabolic method ({\tt PPM}; 
\citealt{Colella and Woodward 1984}) which has been implemented in {\tt PLUTO} code.
The time step $ \Delta t $ is computed using third-order total variation diminishing ({\tt TVD}) Runge Kutta
method ({\tt RK3}) with the CFL number 0.4. In addition, we adopt Harten, Lax, Van Leer approximate 
Riemann solver that restores with the middle contact discontinuity ({\tt HLLC}). Therefore, compared to 
\citealt{Inayoshi et al. 2018, Inayoshi et al. 2019}, we have high-order of accuracy in both space and time. 
Moreover, as we explained in the previous section, we run some models in the presence of thermal conduction.

\subsection{Initial and boundary conditions} \label{sec:initial_boundary_conditions}

As initial condition, we assume a rotating equilibrium torus $ (v_{r} = v_{\theta} = 0) $ with a constant specific 
angular momentum of $ \ell $, embedded in a non-rotating, low-density medium. Starting from the 
momentum equation and considering polytropic equation of state, $ p = A\, \rho^{\gamma} $, 
where $ A $ is a constant, the density distribution of the torus will be determined as,
\begin{equation} \label{initial_torus}
	\frac{\rho_{t}}{\rho_{\infty}} = \left[ 1 + \left(\gamma - 1 \right) \frac{GM}{c_{\rm s, \infty}^{2} r} 
	- \frac{\gamma - 1}{2} \frac{\ell^{2}}{c_{\rm s, \infty}^{2} \varpi^{2}} \right]^{1/(\gamma - 1)},
\end{equation}
where $ \varpi = r \sin \theta $ is the cylindrical radius and the value of adiabatic index is set 
to $ \gamma  = 5/3 $.
When the right-hand side of the above equation
is positive, the density profile is valid. For the region near the rotation axis the right hand-side of 
Equation (\ref{initial_torus}) becomes negative and the ambient medium density is chosen to be 
$ \rho_{a} = 10^{-4} \rho_{\rm crit} $, which is too small to affect our results. The maximum value
of the density, $ \rho_{\rm crit} $, is located at centrifugal radius $ r_{\rm _C} $. 
The initial density distribution is plotted in Figure \ref{contour}.
By defining the constant specific angular momentum as $ \ell = \sqrt{\beta}\, r_{\rm _B} c_{\rm s,\infty} $, where 
$ \beta = r_{\rm _C} / r_{\rm _B} $, the maximum density is evaluated as,
\begin{equation} \label{rho_crit}
	\rho_{\rm crit} = \rho_{\infty} \left[ 1 + \frac{\gamma - 1}{2 \beta} \right]^{1/(\gamma - 1)}. 
\end{equation}

Here we assume $ \beta < 1$ leading to the case of $ r_{\rm s} \ll r_{\rm _C} < r_{\rm _B} $. 
We set $ \beta = 0.1\, r_{\rm _B} $ 
throughout this paper. 
Note that in the absence of viscosity and radiative cooling, the material with $ r_{\rm _C} > r_{\rm _B} $ cannot 
accrete and forms a thick torus near the equator. This thick torus and its formation have been 
a subject of numerous studies (see e.g., \citealt{Papaloizou and Pringle 1984}; 
\citealt{Proga and Begelman 2003a, Proga and Begelman 2003b}). Since the centrifugal radius is very large 
compared to the Schwarzschild radius, the inflowing hot gas always fall to a rotating disk 
before reaching the back hole, due to the outward transport of angular momentum 
mechanism (\citealt{Lynden-Bell and Pringle 1974}; \citealt{Pringle 1981}). To avoid numerical
error, we adopt the density and temperature floor as $ \rho_{\rm floor} = 10^{-4} \rho_{\rm crit} $ and 
$ T_{\rm floor} = 10^{4}\, \rm{K} $ respectively, which is far beyond having any effect to our simulation results.
 
For the boundary conditions, we adopt the outflow boundary condition at the inner and outer 
radial boundaries (e.g. \citealt{Stone and Norman 1992}). We also impose $ v_{r} \le 0 $ at the inner 
boundary which means that inflow of the gas from ghost cells is prohibited. We use the axisymmetric
boundary conditions at both poles $ (\theta = \epsilon, \pi - \epsilon) $.

\section{Results} \label{sec:results}

\begin{deluxetable}{cccc}
\tablenum{1}
\tablecaption{Summary of Models\label{tab:models}}
\tablewidth{0pt}
\tablehead{
\colhead{Model} & \colhead{Thermal Conduction} & \colhead{$ \alpha_{\rm c} $} & \colhead{$ \dot{M}_{\rm acc} $}
}
\decimalcolnumbers
\startdata
A &  NO  & ---  & $ 4.40 \times 10^{-3} $ \\
B & YES & 0.2 & $ 1.51 \times 10^{-2} $ \\
C & YES & 0.5 & $ 2.69 \times 10^{-2} $ \\
\enddata
\tablecomments{The avarage mass accretion rates are in the units of Bondi accretion rate, 
$ \dot{M}_{\rm _B} $, evaluated by Equation \ref{mdot_Bondi}. The mass accretion rates 
are time-averged over $ 2 \le t / t_{\rm orb} \le 4  $. }
\end{deluxetable}

\begin{figure*}[ht!]
\includegraphics[width=\textwidth]{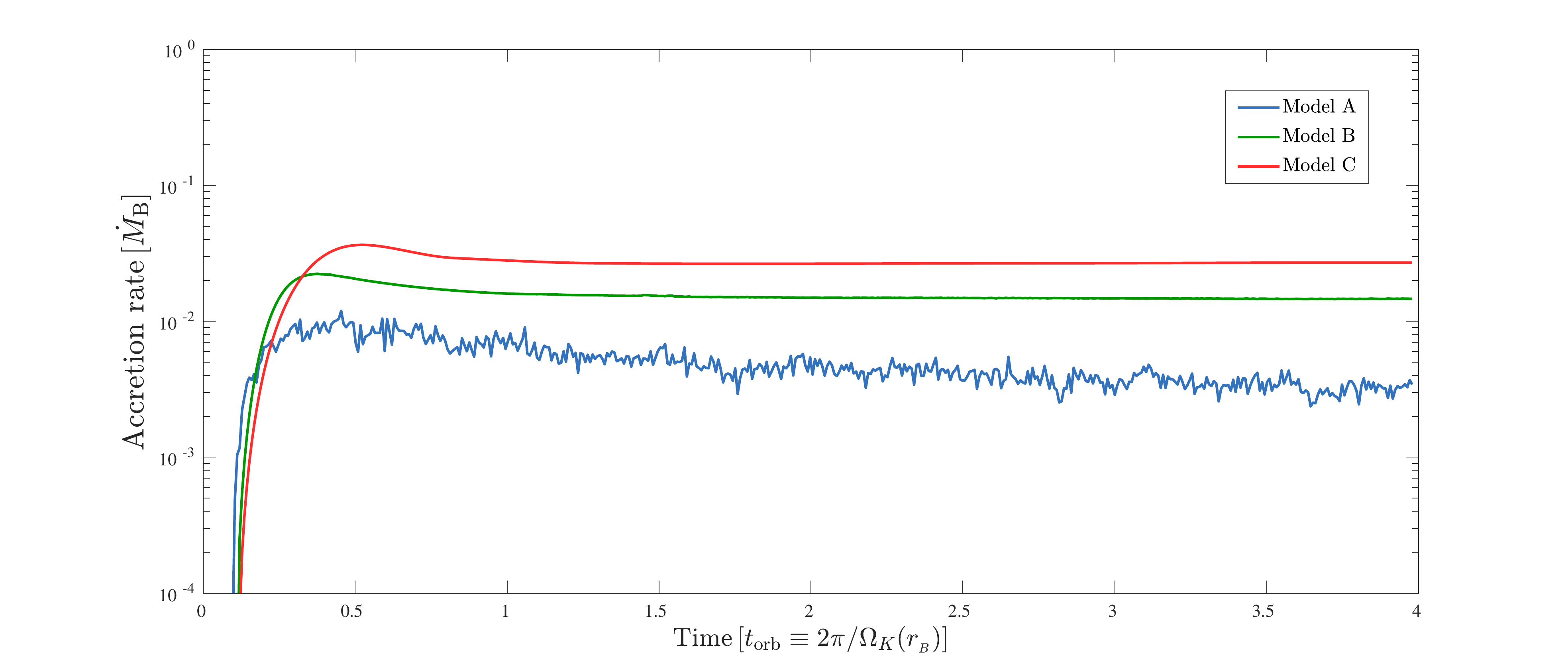}
\centering
\caption{Time evolution of the net accretion rate (in units of the Bondi rate) at $ r_{\rm in} = 10^{-2} r_{\rm _B} $
for Model A (blue), Model B (green) and Model C (red). Here we set the viscosity parameter to be $ \alpha_{_0} = 0.01 $.
\label{mass_accretion}}
\end{figure*}
\subsection{Parameter choices and models} \label{sec:parameter_model}

Before we discuss our simulation results, we introduce the physical units 
and the numerical models that characterize our hot accretion flow simulations  
at large radii. At the initial state, we set the density and temperature of the gas 
at infinity to $ \rho_{\infty} = 10^{-22}\, \rm{g\, cm^{-3}} $ and $ T_{\infty} = 10^{7}\, \rm{K} $, 
respectively. This temperature is equivalent to the sound speed of 
$ c_{\rm s, \infty} = \sqrt{\gamma k_{\rm B} T_{\infty} / (\mu m_{\rm p})} = 4.7 \times 10^{7} \rm {cm\ s^{-1}} $
with $ \mu = 0.62 $. In our units, we scale all the spatial scales with $ r_{\rm _B} $, 
velocities with $ c_{\rm s, \infty} $, and density with the density at infinity $ \rho_{\infty} $.
Therefore, units of time, pressure $ p $, and the kinematic viscosity coefficient $ \nu $ 
will be $ r_{\rm _B} / c_{\rm s, \infty} $,  $ \rho_{\infty} c_{\rm s, \infty}^{2} $ and 
$ c_{\rm s, \infty}  r_{\rm _B} $, respectively\footnote{In the absence of cooling terms, 
equations (\ref{eq:continuity})-(\ref{eq:energy}) are independent of 
the density normalization. The reason is that both $ p $ and $ \bm{\sigma}$ are proportional 
to the density as $ p \propto \rho $ and $ \bm{\sigma} \propto \rho $, provided that the kinematic viscosity 
coefficient $ \nu $ is not a function of density. Therefore, these results can be scaled to a wide 
range of systems--from hot accretion flows around stellar mass black-holes, such as X-ray binaries
(XRBs), to those around SMBHs. So, the simulation without cooling, such a hot accretion flow 
models is invariant to any change in the density normalization and equivalently to the mass accretion rate.}. 
Unless stated otherwise, time-scales 
are evaluated in terms of the orbital time scale of a test particle at Bondi radius which is given by
$ t_{\rm orb} = 2 \pi / \Omega_{\rm _K} (r_{\rm _B}) $. Bondi accretion rate is well defined 
and standard reference of the accretion rate from the Bondi radius which can be written as 
\begin{equation} \label{mdot_Bondi}
	\dot{M}_{\rm B} = 4 \pi \lambda(\gamma) \rho_{\infty} \frac{G^{2} M^{2}}{c_{\rm s, \infty}^{3}}, 
\end{equation}
where $ \lambda(\gamma) = 1/4 $ for $ \gamma = 5/3 $.
We should note that the accretion parameter $ \lambda $ can strongly depend 
on the galaxy potential at very large distances from the central BH, and therefore 
the Bondi accretion rate will be much larger than the expected for assigned density and 
temperature at infinity (e.g., see \citealt{Ciotti and Pellegrini 2017, Ciotti and Pellegrini 2018}; 
\citealt{Mancino et al. 2022}).  We normalize mass accretion rates with Bondi accretion 
rate throughout this paper. The viscous parameter is set to $ \alpha_{_0} = 0.01 $.  

In this study, we perform hydrodynamical simulations of hot accretion 
flow at large radii. In all simulations presented here, the viscosity is modeled 
base on Equations (\ref{sigma_ij})-(\ref{alpha_vis}) 
and we ignored the radiative losses. In model A we simulate hot accretion flow without 
thermal conduction and investigate the existence of wind. Moreover, we carry out two models, 
i.e., models B and C, to study the effects of thermal conduction on wind. The conductivity 
coefficient of model B and C is considered as $ \alpha_c = 0.2 $ and $ 0.5 $, respectively.
Details of the simulation models are tabulated in Table 1. The last column of 
Table \ref{tab:models} represents the time-averaged mass accretion rate through the inner 
boundary in the units of Bondi accretion rate.

\begin{figure*}[ht!]
\includegraphics[width=\textwidth]{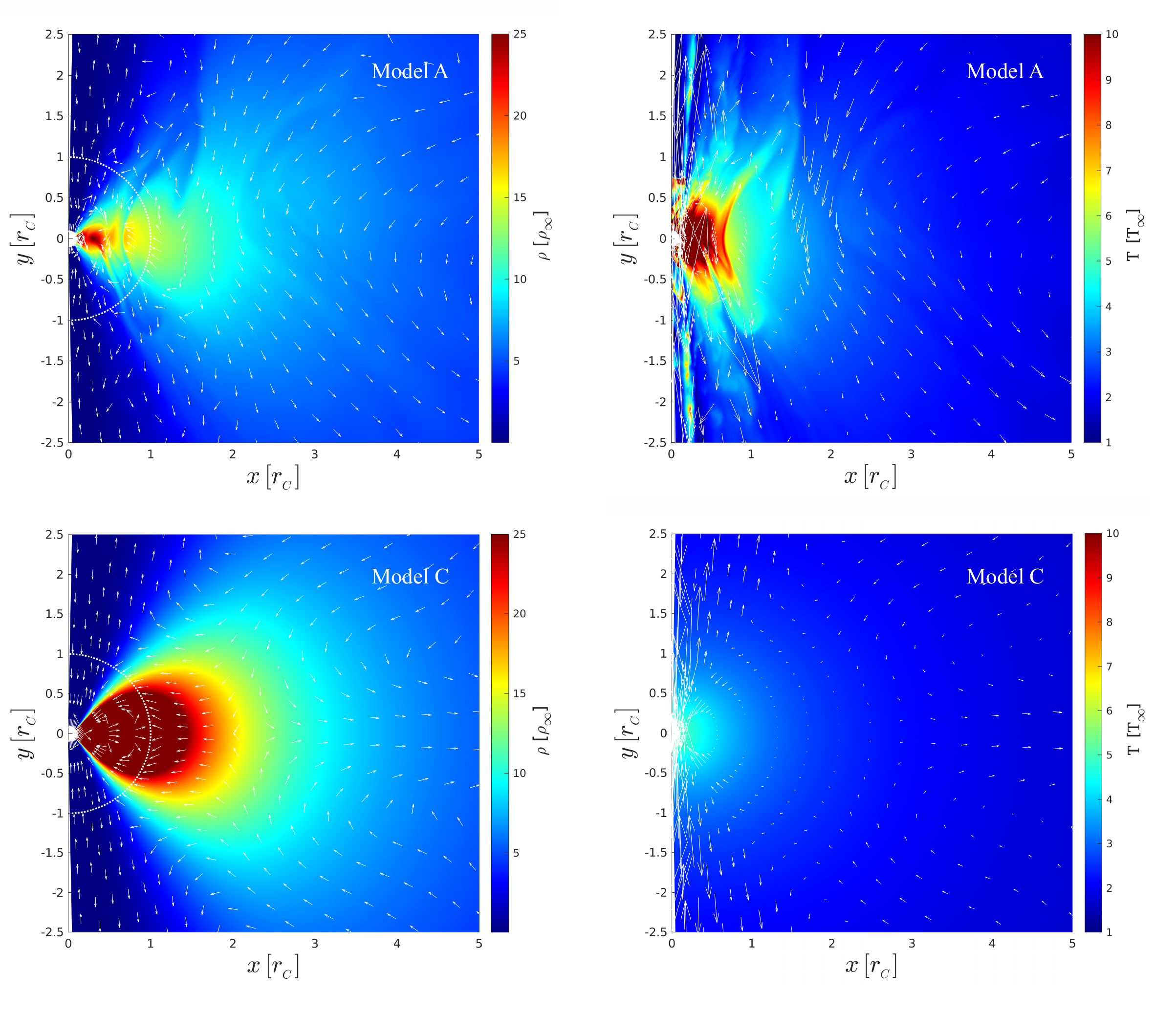}
\centering
\caption{Snapshots of contours of gas density (left-hand column) and temperature 
(right-hand column) correspond to models A and C, respectively. The elapsed time 
is set to $ t = 4 t_{\rm orb} $. The densities and temperatures are over-plotted with the 
poloidal velocity $ \bm{v}_{\rm p} = v_{r} \hat{\bm{r}} + v_{\theta} \hat{\bm{\theta}} $. 
Arrows on the density profiles show only the direction of velocity $ (\bm{v}_{\rm p} / \left| \bm{v}_{\rm p} \right|) $,
while the arrows on the temperature profiles indicate both the magnitude and direction of the poloidal velocity. 
The dotted white curves in the left panels show the location of the centrifugal radius $ r_{\rm _C} = 0.1 r_{\rm _B} $. 
\label{contours}}
\end{figure*}

\subsection{Mass inflow rate} \label{mass_inflow_rate}

Following \citealt{Stone et al. 1999}, the mass inflow and outflow rates, $ \dot{M}_{\rm in} $ 
and $ \dot{M}_{\rm out} $, will be defined as, 
\begin{equation} \label{eq:mdot_in}
	 \dot{M}_{\rm in}  = 2 \pi r^{2} \int_{0}^{\pi} \rho \, \mathrm{min}(v_{r}, 0) \sin \theta d\theta,
\end{equation} 

\begin{equation} \label{eq:mdot_out}
	 \dot{M}_{\rm out} = 2 \pi r^{2} \int_{0}^{\pi} \rho \, \mathrm{max}(v_{r}, 0) \sin \theta d\theta.
\end{equation}

The total mass outflow rate calculated by equation (\ref{eq:mdot_out}) is includes real outflow 
and outward moving portion of turbulent eddies, since this consists of all of the fluid with positive 
velocity. In the following, we use wind to denote real outflow.
\begin{figure*}[ht!]
\centering
\includegraphics[width=0.68\textwidth]{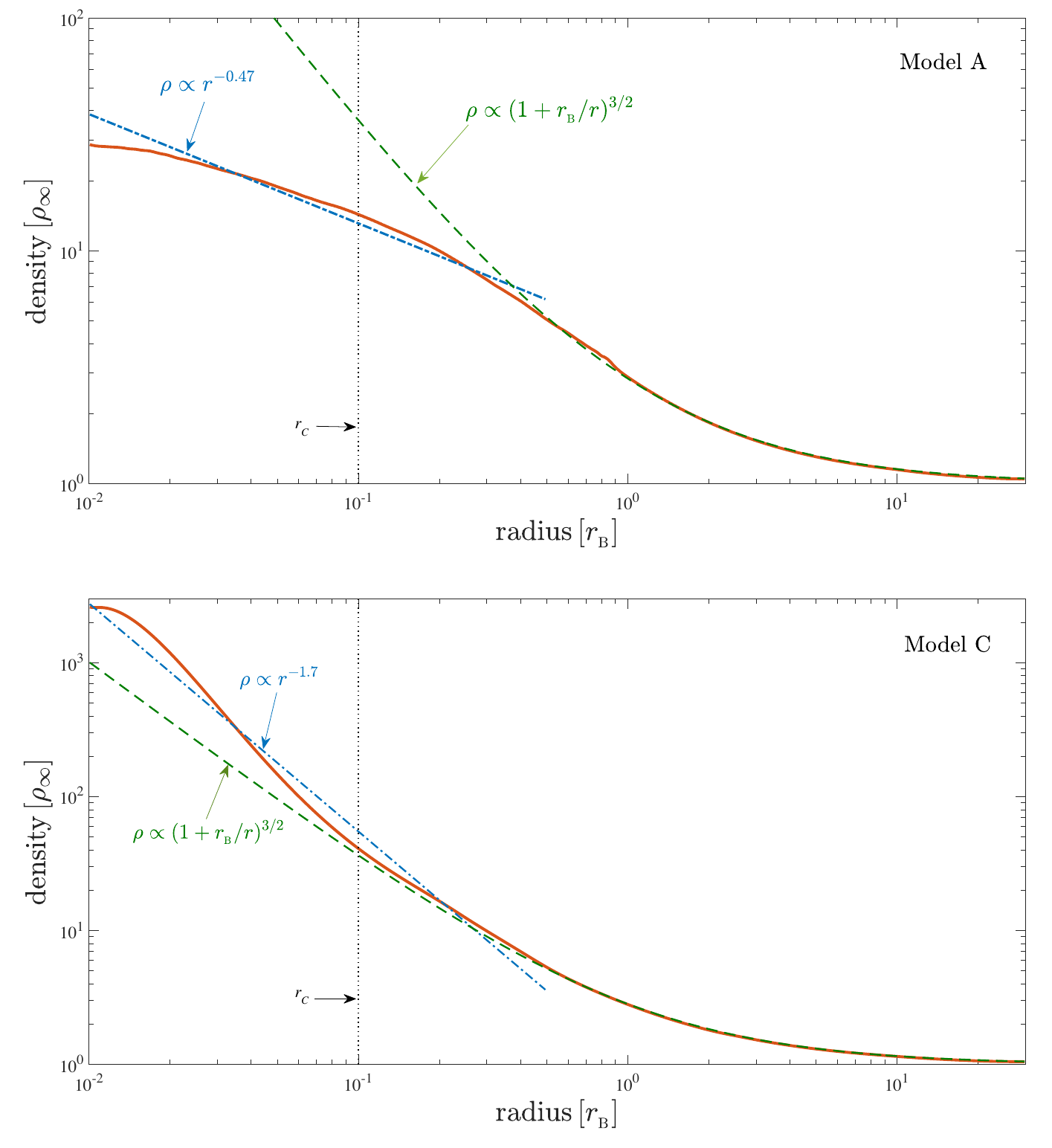}
\caption{Radial profile of the gas density along the equatorial plane for Models A (top panel) and C (bottom panel). 
The density profiles are time-avarge over $ 2 \le t / t_{\rm orb} \le 4  $ and angle-average over 
$ 80^{\circ}  \le \theta \le 110^{\circ} $. The dotted lines show the location of circularization radius, $ r_{\rm _C} = 0.1 r_{\rm _B} $. 
At the region of $ r \ge 2 r_{\rm _C} $, the density profiles of both models approximately follow $ \rho \propto (1 + r_{\rm _B} / r)^{3/2} $,
(green dashed lines). In the inner region, the radial density profile of model A follows $ \rho \propto r^{-0.47} $, while for model C, 
in the presence of thermal conduction, it follows $ \rho \propto r^{-1.7} $ (blue dash-dotted lines).
\label{fig:density_time_avarage}}
\end{figure*} 
The net mass accretion rate can be written as,
\begin{equation} \label{eq:mdot_acc}
	 \dot{M}_{\rm acc}  = 2 \pi r^{2} \int_{0}^{\pi} \rho \, v_{r} \sin \theta d\theta,
\end{equation}
which is $  \dot{M}_{\rm acc}  = \dot{M}_{\rm in} +  \dot{M}_{\rm out} $. Figure \ref{mass_accretion} 
shows the time evolution of the angle-integrated mass accretion rate $ \dot{M}_{\rm acc} $ at
$ r = 0.01 r_{\rm _B} $ (i.e., the inner boundary) for Model A (blue), Model B (green) and Model C
(red). We normalize the simulation time by the orbital time-scale $ t_{\rm orb} $ at the Bondi 
radius and the mass accretion rate by Bondi accretion rate $ \dot{M}_{\rm _B} $. 
Since the angular momentum of the gas is set equal to the Keplerian angular momentum 
at $ r_{\rm _C}= 0.1 r_{\rm _B} $, it is natural that the gas at the outer part tends to accumulate 
around this region because of the black hole gravity. Moreover, due to the form of viscosity 
(see equation (\ref{alpha_vis})), this process starts to work at the region of $ r < r_{\rm _C} $, 
so that the angular momentum of the gas can be transported outward, which drives inflow 
gas motion in a quasi-steady fashion.

Our runs have been evolved around $ 4 $ orbits at Bondi radius. From this figure we can see that for 
all models, the mass accretion rate rapidly increases until $ t \leq 0.5 t_{\rm orb} $. Then the quasi-steady 
accretion phase starts at the range of $ 0.5 \, t_{\rm orb} < t \le 4 \, t_{\rm orb} $ and all the physical quantities
become nearly constant. The fourth column of Table \ref{tab:models} also shows the corresponding 
time-averaged mass accretion rate over $ 0.5 \, t_{\rm orb} < t \le 4 \, t_{\rm orb} $.  
For Model A, the accretion rate has fluctuations at $ t > 0.5 \, t_{\rm orb} $ and oscillates with very 
small amplitude around its mean value $ \dot{m} = \dot{M}_{\rm acc} / \dot{M}_{\rm _B} \simeq 4.40 \times 10^{-3} $. 
The behavior of the accretion rates in model B and C, where the thermal conduction is presented, 
is similar and greater than model A. In fact, the time-averaged values of Model B and C in 
the quasi steady-state case are $ \dot{m} = 1.51 \times 10^{-2} $ and $ \dot{m} = 2.69 \times 10^{-2} $ respectively 
(see the column 4 of Table \ref{tab:models}). This result shows that thermal conduction increases the 
mass accretion rate by one order of magnitude from the case without conduction. 
From Figure 2, it appears that model A (model with no thermal conditions) 
is noisier than models B and C where the thermal conditions is presented. 
This is due to the rapid time variation of poloidal velocity and gas density in 
model A than models B and C.  The main reason is as follows: in model A, 
where the thermal condition is not included, the accretion flow is quite turbulent 
and convective motions result in the turbulent nature of the accretion flow. 
On the other hand, for models with thermal conduction, model B and C, 
the accretion flow becomes quite laminar which means the convective 
motion is not important and significantly supressed.

\subsection{Our fiducial models}

We choose model A (without thermal conduction) and model C (with thermal conduction) as 
our fiducial models in this paper. Figure \ref{contours} shows the snapshot of the gas 
density (left-hand column) and temperature (right-hand column) corresponding to the 
models A and C. The elapsed time is set to $ t = 4 t_{\rm orb} $. 
The densities and temperatures are over-plotted with the 
poloidal velocity $ \bm{v}_{\rm p} = v_{r} \hat{\bm{r}} + v_{\theta} \hat{\bm{\theta}} $. 
Arrows on the density profiles show only the direction of velocity $ (\bm{v}_{\rm p} / \left| \bm{v}_{\rm p} \right|) $,
while the arrows on the temperature profiles indicate both the magnitude and direction of the poloidal velocity. 
The dotted white curves in the left panels show the location of the centrifugal radius $ r_{\rm _C} = 0.1 r_{\rm _B} $. 
From the density profile of both models, it is clear that the maximum density 
region is located inside $ r_{\rm min} < r < 2 r_{\rm _C} $. 
The reason is that the angular momentum $ \ell $ begins to be transported outwards due to 
definition of the viscosity parameter in our models. Equation (\ref{alpha_vis}) shows that viscosity 
highly depends on the threshold of the density, $ \rho_{\rm crit} $, above which the viscosity will be 
turned on. The density around the midplane has increased in model C, with thermal conduction. 
To know how thermal conduction affects the temperature of the flow, we should compare the right panels 
of Figure \ref{contours}. In both top and bottom right panels of this figure, the gas temperature is 
increasing toward the centre due to the compression heating of the black hole gravity and dissipation
heating produced by viscosity.  It is also clear that in the model without thermal conduction, the 
temperature in the dense core of the accretion flow becomes higher than $ T > 10^{8}\, K $ while the 
maximum temperature in model C with thermal conduction is about $ T \sim 5-6 \times 10^{7} $. 
Temperature decrease in model C will result in the density increase. This is because when the temperature 
increase in the vertical direction, the pressure will support the disk. Consequently, if the temperature 
decreases, the density will increase to sustain the fix pressure. In model C, the temperature is also 
distributed homogeneously around the central black hole. 

To show that why the net mass accretion rate increases in models with thermal conduction, both the
density and temperature profiles are overplotted with the direction and magnitude of the poloidal velocity.
From the right column of Figure \ref{contours} we see that in model C most fraction of 
the gas is inflowing from inside the Bondi radius and is directed towards the central black hole. On the other hand,
for model A we observe strong wind (long arrows) at the disk surface and also inflow and outflow 
motions in circulation at this inner region. From the arrows in the left column we can see that in both models, 
very close to the rotation axis, the gas is inflowing towards the central black hole. Moreover,
in model A the mirror symmetry of the hot accretion flow cross the midplane is broken. In model C, where
the thermal conduction is presented, the accretion disk is almost symmetric above and below the equatorial plane.

The density profiles as well as magnitude of the poloidal velocity clearly show the reason 
why the mass accretion rate in the presence of thermal conduction has increased one order 
of magnitude compared to the case without conduction (see, Equation (\ref{eq:mdot_acc}) for more details 
about the dependency of net accretion rate to the density and radial velocity).

\begin{figure*}[ht!]
\includegraphics[width=0.9\textwidth]{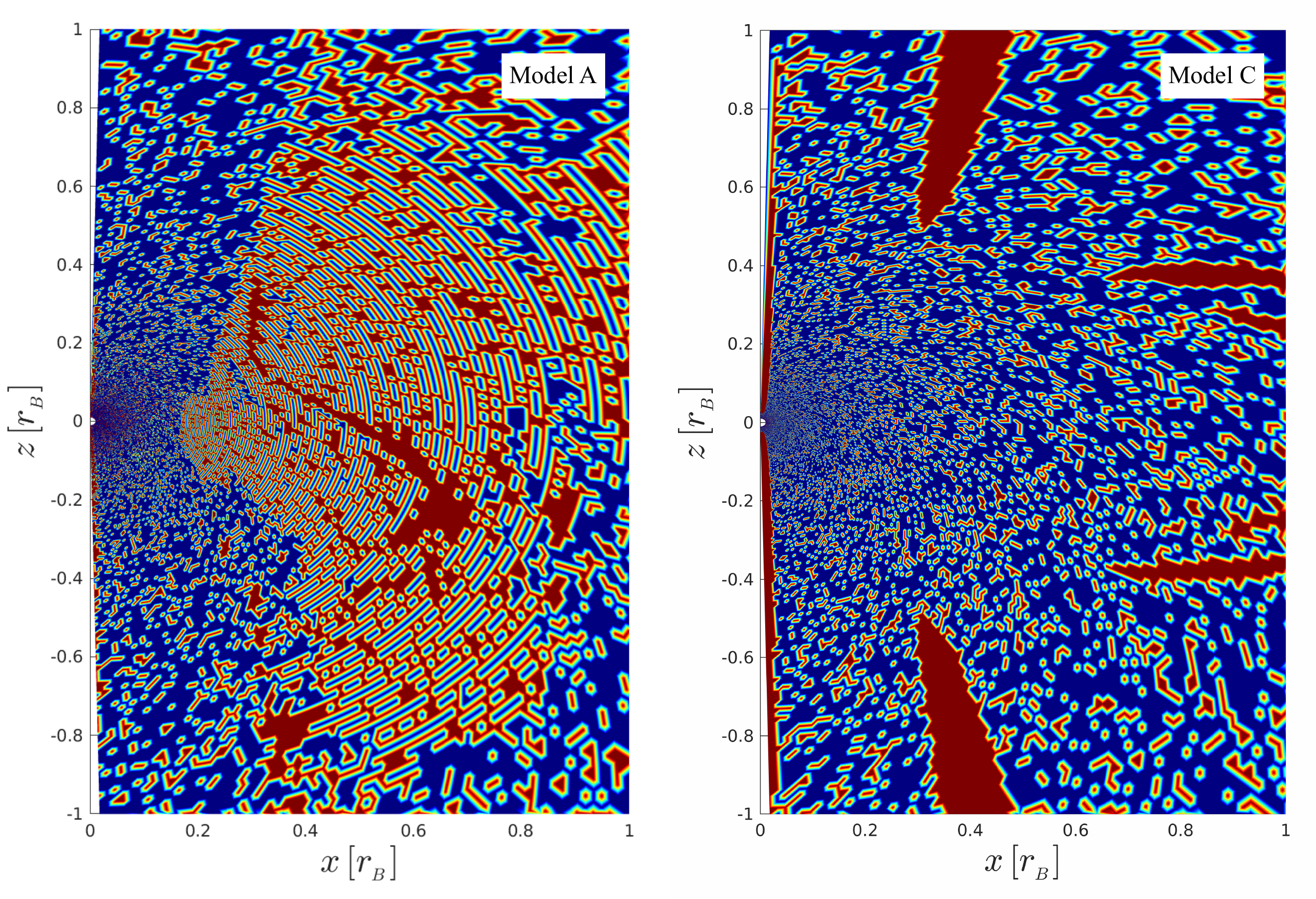}
\centering
\caption{Convective stability analysis of models A (left panel) and C (right panel).
The results are obtained according to Equations (\ref{Hoiland_1}) and (\ref{Hoiland_2}) 
based on simulation data time-averaged over $ 2 \le t / t_{\rm orb} \le 4  $. 
The red color denotes unstable regions. \label{convective_stability}}
\end{figure*}

Figure \ref{fig:density_time_avarage} shows the radial profile of gas density in the equatorial plane
for model A (top panel) and C (bottom panel) in order to understand the properties of hot accretion flow
with/without thermal conduction. The density profiles are time-averged over $ 2 \le t / t_{\rm orb} \le 4 $,  
and angle-average over $ 80^{\circ}  \le \theta \le 110^{\circ} $. The dotted lines show the location of 
circularization radius, $ r_{\rm _C} = 0.1 r_{\rm _B} $. From this figure we can see that at the outer region 
accretion flow, $ r \ge 2 r_{\rm _C} $, the density profiles of both models perfectly follow $ \rho \propto (1 + r_{\rm _B} / r)^{3/2} $,
(see green dashed lines). For the inner region, the radial density profile of model A fit well with 
$ \rho \propto r^{-0.47} $, while for model C, in the presence of thermal conduction, it follows 
$ \rho \propto r^{-1.7} $ (see blue dash-dotted lines). This implies that when the thermal conduction 
is not considered in the flow, the profile of the density has a smoother trend at the inner region 
as shown in this figure. We emphasize here that, based on the radial profile of the density 
at the equatorial plane, \citealt{Inayoshi et al. 2018} argued that this profile is the consist of two components: 
(1) a rotational equilibrium solution at the outer region ($ r > 2 r_{\rm _C} $) which follows 
$ \rho \propto (1 + r_{\rm _B} / r )^{3/2} $, and (2) a convection dominated regions at 
the inner region ($ r < 2 r_{\rm _C} $) follows $ \rho \propto r^{-1/2} $. More precisely, based on 
the density profile, they argued that decrease of mass accretion rate inward is because of convective 
motion. Therefore they concluded that the inner region agrees well with the CDAF solution and there
is no wind. In the following subsection we investigate the convective stability of the 
hot accretion flow with/without thermal conduction. 

\subsection{Convective Stability}

Numerical HD simulations of hot accretion flow have been found that the flows are convectively unstable 
(see i.e., \citealt{Stone et al. 1999}; \citealt{Igumenshchev and Abramowicz 1999, Igumenshchev and Abramowicz 2000}; 
\citealt{Yuan and Bu 2010}; \citealt{Bu et al. 2016a}). 
This also have been suggested by one-dimensional self-similar solutions of \citealt{Narayan and Yi 1994}. The main
physical reason is as follow, due to the viscous dissipative heating and negligible radiative loss, the entropy 
of the hot accretion flow increases inward. In this subsection we investigate the convective stability 
of hot accretion flows at large radii based on our numerical simulation results of models A and C. 
To treat the convective stability of the flow, the well-known Solberg-H\o iland criterions 
in cylindrical coordinates $ (\varpi, \phi, z) $ will be adopted (e.g., \citealt{Tassoul 2000}). 
If the hot accretion is convectively stable, both of the two following criteria shall be positive:

\begin{equation} \label{Hoiland_1}
	- \frac{1}{\gamma \rho} \bm{\nabla} p \cdot \bm{\nabla} s + \frac{1}{\varpi^{3}} \frac{\partial \ell^{2}}{\partial \varpi}  \geq 0,
\end{equation}

\begin{equation} \label{Hoiland_2}
	- \frac{\partial p}{\partial z} \left( \frac{\partial \ell^{2}}{\partial \varpi} 
	\frac{\partial s}{\partial z} -  \frac{\partial \ell^{2}}{\partial z} \frac{\partial s}{\partial \varpi} \right) \geq 0,
\end{equation}
where $ s  = \ln(p/ \rho^{\gamma}) $ is the entropy. 
We adopt the following transformations to find the angular dependency of two Solberg-H\o iland criteria
in spherical coordinates,

\begin{align} \label{eq:transformations}
	\frac{\partial}{\partial \varpi} = \sin \theta \frac{\partial}{\partial r} + \frac{\cos \theta}{r} \frac{\partial}{\partial \theta}, \\ 
	\frac{\partial}{\partial z} = \cos \theta \frac{\partial}{\partial r} - \frac{\sin \theta}{r} \frac{\partial}{\partial \theta}.
\end{align}

\begin{figure*}[ht!]
\includegraphics[width=0.7\textwidth]{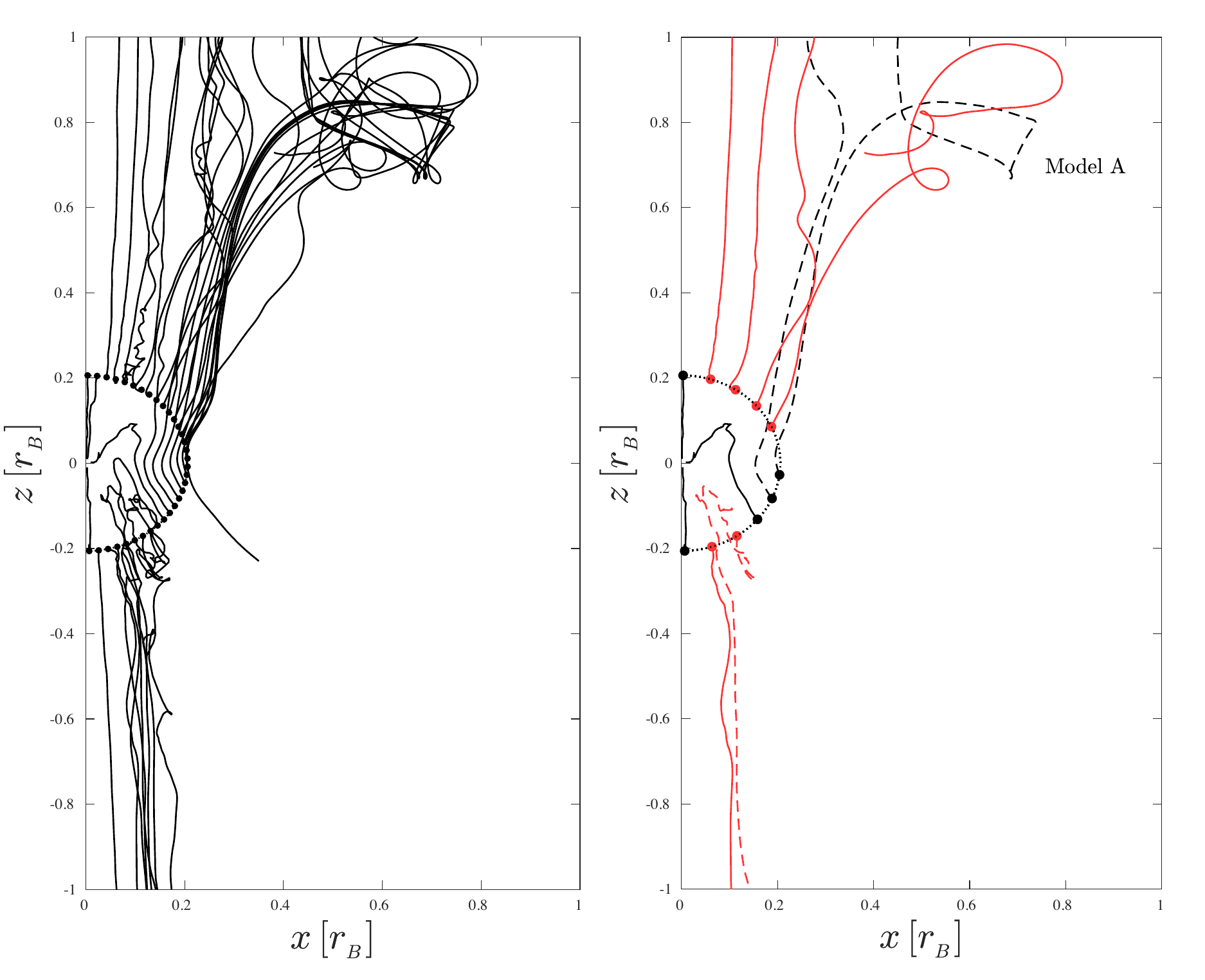}
\centering
\caption{Various types of trajectory of test particles for model A. The dots located at 
$ r = 0.2 r_{\rm _B} $ are starting points of the particles. In the left panel we select 34
particles in diferent $ \theta $ angles and significant winds are clearly present. In the right
panel, the red lines denote outflow, while black ones are for inflow. See subsection 
\ref{sec:trajectory_method} for details. \label{fig:trajectory_modelA}}
\end{figure*}

The convective stability analysis of models A and C are shown in Figure \ref{convective_stability}. 
The results are obtained according to Equations (\ref{Hoiland_1}) and (\ref{Hoiland_2}) 
based on simulation data time-averaged over $ 2 \le t / t_{\rm orb} \le 4 $. 
The red color denotes unstable regions. The result indicates that in both models the 
convectively unstable regions exist. The physical reason is the same as we 
explained before: during the accretion process, in the absence of radiative cooling, 
due to the viscous dissipated heating the entropy of the gas increases inward. 
Moreover, although in both models convectively unstable regions exist, for model 
C when the thermal conduction is included (right panel), the stable regions are 
larger and the convection motions are suppressed. 

Previous works such as \citealt{Yuan et al. 2012a} showed that strong winds are produced in HD hot accretion
flows around black holes. They found that the driving force for wind production in such a system is mainly
buoyant force associated with the convection instability. In the next subsection, following \citealt{Yuan et al. 2015},
we use trajectory particle method to study whether winds exist at such large radii.

\subsection{Trajectory Method} \label{sec:trajectory_method}

In the turbulent flow the existing turbulent eddies move outwards and consequently 
have positive radial velocities. In principle, they are not real outflow and are only portion of
turbulent motions. On the other hand we have real outflow, where the flow go outward and escape 
to the large radii. The real outflow is called wind. One technique that we adopt to characterize 
and identify wind based on our mesh-based simulations is ''trajectory test particles'' which are 
passively advected with the flow, and thereby track its Lagrangian evolution, allowing the 
thermodynamical history of individual fluid elements to be recorded. This technique is named 
La‌grangian particle tracking method and has been used in several different astrophysical simulations. 
We utilize the trajectory method to distinguish between turbulent outflow and real outflow (see, \citealt{Yuan et al. 2015}). 
The difference between outflow and real outflow is that the outflow might be turbulent outflow 
which particles rejoin the flow after flowing outward. The real outflow is considered for the particles 
that go outward and escape the outer boundary of the simulation domain. As we mentioned, trajectory 
is related to the Lagrangian description of the fluid, and obtained by following the motion of fluid elements 
at consecutive times. The superiority of the trajectory method to the streamlines is that it is used for the 
turbulent motion such as accretion flow. The trajectory is analogous to the streamline just when the 
flow has steady motion. 

\begin{figure*}[ht!]
\includegraphics[width=0.7\textwidth]{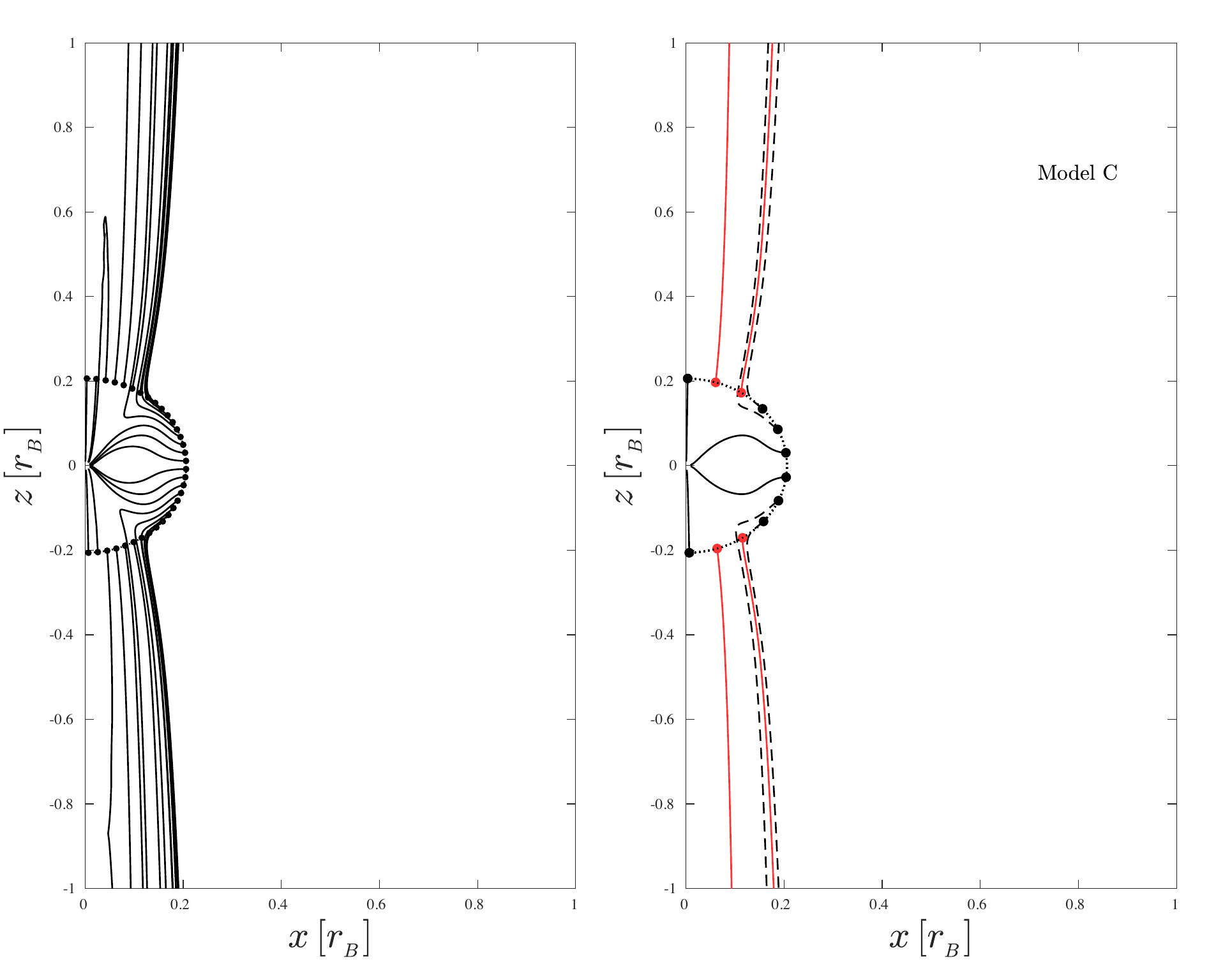}
\centering
\caption{Same as figure \ref{fig:trajectory_modelA}, but for model C. \label{fig:trajectory_modelC}}
\end{figure*}

As it is stated before, \citealt{Inayoshi et al. 2018} performed similar simulations in the absence of
thermal conduction and based on the radial profile of density, they argued that there exists no outflow. 
More precisely, they claimed that convective motions results in inward decrease of mass accretion rate. 
Here, using trajectory method, we will consider a set of virtual test particles in the simulation domain 
at different radii to extensively study the wind. Following \citealt{Yuan et al. 2015}, We use ``outflow'' to 
describe any flow with a positive radial velocity i.e., $ v_{r} > 0 $, flowing outward. This can be both 
``turbulent outflow'' and ``real outflow''. The main difference between them is that in the turbulent case 
the test particle will return and join the accretion flow after flowing outward for some distance. On the 
other hand, in the case of real outflow the test particle continues to flow outward and eventually 
escapes the outer boundary of our computational domain. 

\begin{figure*}[ht!]
\plottwo{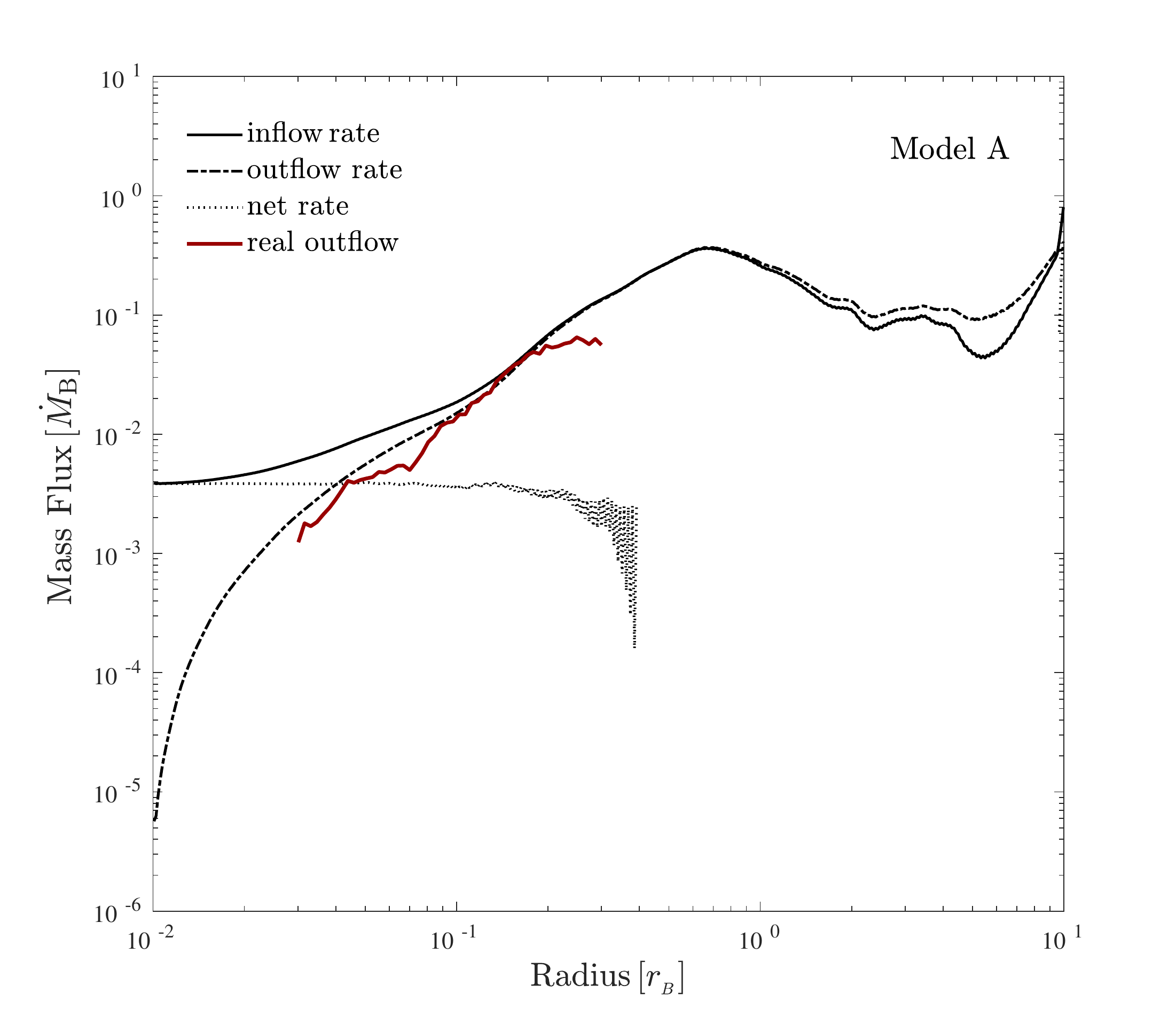}{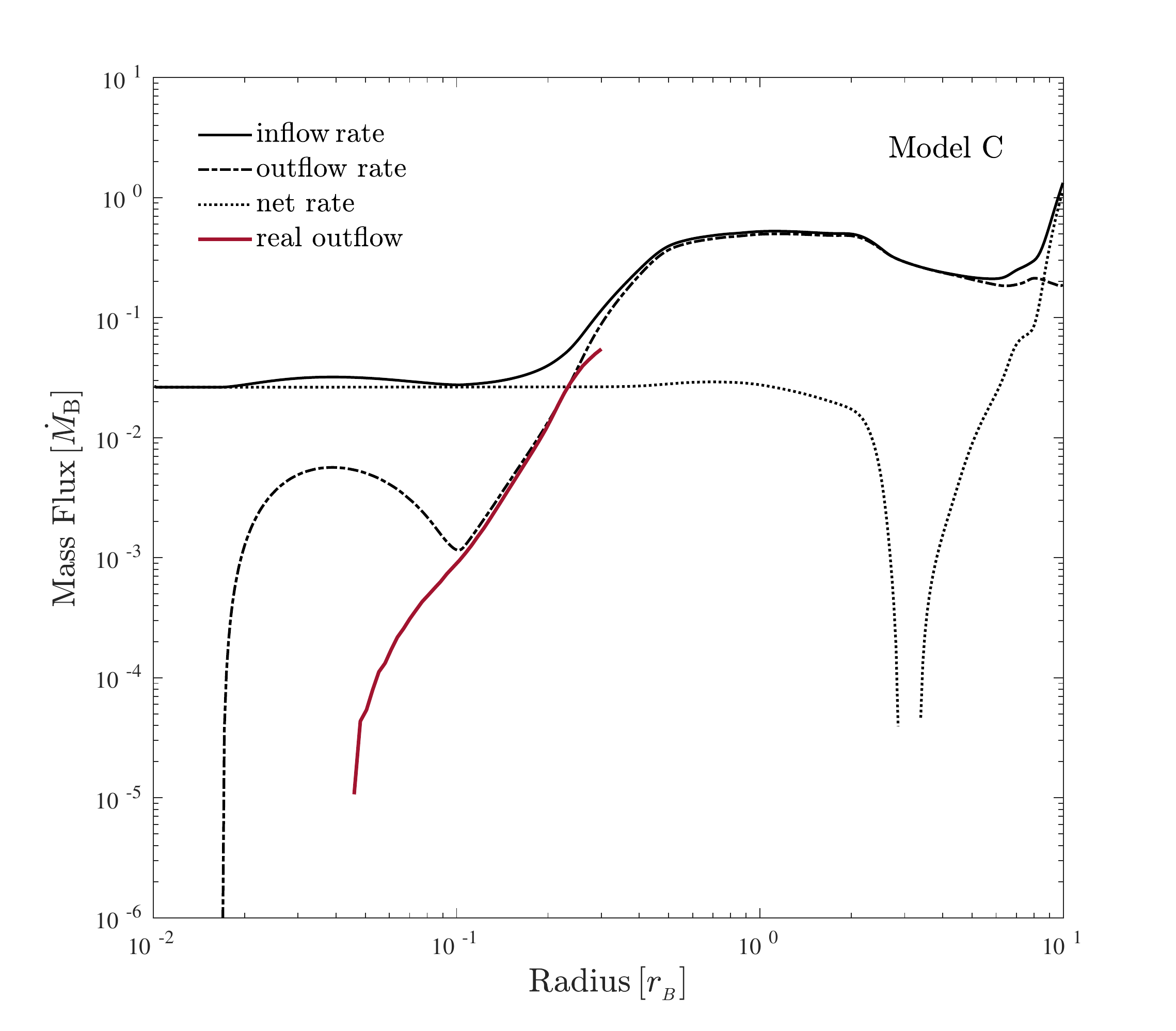}
\centering
\caption{Radial profile of the time-averaged mass inflow rate (solid black line), outflow rate (dash-dotted line), 
net rate (dotted line), and real outflow (solid red line) for models A (left panel) and C (right panel). 
The results are time-averaged over $ 2 t_{\rm orb} \le t \le 4 t_{\rm orb} $. We do the trajectory analysis inside 
$ 0.3 r_{\rm _B} $, since the quasi-steady state solutions achieved in this region. \label{mass_flux}}
\end{figure*}

By adopting the trajectory method, we can discriminate which particles are real outflow and
which ones are turbulent motions. The left panel of Figure \ref{fig:trajectory_modelA} shows
the trajectory of 34 test particles staring from $ r = 0.2 r_{\rm _B} $ in model A (without thermal 
conduction). From this figure we can clearly see the real wind trajectories, i.e., the particles 
that extend from $ r = 0.2 r_{\rm _B} $ to the large radii above the Bondi radius would never cross
$ r = 0.2 r_{\rm _B} $ twice. Note here that distinguishing the types of characteristic particle 
trajectories is crucial for calculating the mass fluxes of the real outflow correctly. Therefore, 
in the right panel of Figure \ref{fig:trajectory_modelA} we distinguish inflow and outflow with 
black and red lines, respectively. Here, the red solid lines show the real outflow, where the 
particles keep moving outward and never cross the radius $ r = 0.2 r_{\rm _B} $ again. 
Moreover the red dashed line represents the turbulent outflow, where the particle first move 
outward but will return and cross the radius $ r = 0.2 r_{\rm _B} $ 
during its motion, and eventually move outward. In this figure, the black solid lines represent real inflow
while the black dashed lines are turbulent inflow.

Figure \ref{fig:trajectory_modelC} is the same as figure \ref{fig:trajectory_modelA}, but for 
model C in the presence of thermal conduction. Based on this figure we can see that in the 
region around equatorial plane, the particles are inflow particles which move 
toward the central black hole, while at the high latitudes we have more outflow motions. 
Compared to the model A, in model C, the trajectory particles have a symmetrical traces  
above and below the equator. In addition, in both models the particles very close to the rotation axis move
towards the centre as inflow particles.

After obtaining the trajectories, we calculate the mass fluxes in the last part of this section. 
Figure \ref{mass_flux} shows the radial profile of the time-averaged (from $ t = 2 $ 
to 4 orbits) mass inflow rate (solid black line), outflow rate (dash-dotted line), the net rate (dotted line), 
calculated from Equations (\ref{eq:mdot_in})-(\ref{eq:mdot_acc}). The red lines denote the 
mass flux of the real outflow evaluated by trajectory method. We do the trajectory 
analysis inside $ 0.3 r_{\rm _B} $, since the quasi-steady state solutions are achieved in this region.
This figure clearly shows that for both models, the wind exist inside the Bondi radius. 
The inflow and outflow rates decrease toward the central black hole. 
The inflow profile in model C is more flattened with respect to model A.
More specifically, in the region $ 0.01 < r < 0.1 $ of model C, the prominence of the inflow over the outflow
is due to the existence of thermal conduction. The reason is that conduction can take away the release 
gravitational energy of the accretion flow. In this case it is not necessary to have a strong wind to take a way the
release gravitational energy for the accretion process to occur. In model A the mass outflow rate increases 
from larger radii, and reaches to maximum amount at $ 0.8\, r_{\rm _B} $ compared with model C. 
Even more, in model A the real outflow starts to begin at smaller radius around $ 0.03\, r_{\rm _B} $ 
while it starts around $ 0.04 r_{\rm _B} $ in model C. 

In addition, to quantitatively investigate the wind, we calculated the ratio of mass flux of winds to 
the total outflow rate at $ r = 0.2 r_{\rm _B} $.  
This ratio for model A and C is $ 84 \% $ and $ 78 \% $ respectively.  
Our quantitative calculation of the ratio of mass flux of winds to the total outflow rate 
based on trajectory method clearly confirms that the wind exists in hot accretion 
flow and can reach to large radii. Although in previous study of \citealt{Inayoshi et al. 2018} 
they did not find wind in their solution, our results  show that very effective wind can 
be produced inside the Bondi radius. Therefore, we conclude that the decrease of 
mass accretion rate inward is due to the wind rather than the convection at such large radii. 

\section{Summary} \label{sec:summary}

Numerical HD and MHD simulations of hot accretion flow around black hole show that strong wind must
be present in such a system. For instance, \citealt{Yuan et al. 2015} found that the mass flux of wind follows 
$ \dot{M}_{\rm wind} = \dot{M}_{\rm _{BH}} (r / 20 r_{\rm s}) $. Subsequently, the main question is 
how far the wind can be produced? In order to answer this question, we study hot accretion flow
on a SMBH with two-dimensional hydrodynamical simulations. The accretion flow is considered 
to be axisymmetric. We only take into account the gravity of the central black hole and neglect 
the nuclear stars gravity. Moreover, the radiative losses and relevant AGN feedback are not 
considered in this study. In our hydrodynamical simulations, we adopt $ \alpha $ prescription of viscosity to mimic the effect 
of the magnetic stress. In some of our models, we include thermal conduction and compare the results 
with the case without conduction. The runs have been evolved around 4 orbits at Bondi radius. For all of
our models, the mass accretion rate from the inner boundary of our computational domain becomes 
almost constant at the range of $ 0.5 t_{\rm orb} < t \le 4 t_{\rm orb} $ which indicates that the flow reaches 
to the steady state. For our model without thermal conduction, the mass accretion rate has some oscillation at 
$ t > 0.5 t_{\rm orb} $ with very small amplitude around its mean value, while the accretion rates in 
models with thermal conduction are almost constant and greater than model A (a model with thermal 
conduction). Furthermore, the density around the midplane has increased in models incorporating thermal 
conduction and a thick and hot accretion flow forms. 

We investigate the convective stability of the hot accretion flow with/without thermal conduction
at large radii based on our numerical simulation results. Our results show that for both models 
the disk is convectively unstable. Although both models are convectively unstable, 
thermal conduction could slightly decrease the instability (see Figure \ref{convective_stability}). 
In previous simulation of \citealt{Inayoshi et al. 2018}, they found a global steady accretion solution with two 
distinguished regimes: (1) the outer rotational equilibrium region around Bondi radius follows
 the density profile of $ \rho \propto (1 + r_{\rm _B} / r)^{3/2} $ and (2) the inner solution where 
the geometrically thick torus follows $ \rho \propto r^{-1/2} $. Based upon the density profile of 
inner part, they argued that the physical properties of this region are consistent with the 
convection-dominated accretion flows (CDAFs) rather than Adiabatic Inflow-Outflow Solutions
(ADIOS). In order to understand the properties of hot accretion flow with/without thermal 
conduction, the density profiles at the equatorial plane have been checked. 
We found that at the outer region of accretion flow, $ r \ge 2 r_{\rm _C} $, the density 
profiles of all models perfectly follow $ \rho \propto (1 + r_{\rm _B} / r)^{3/2} $. 
In the inner region, the radial density profile of model without thermal conduction fits well with 
$ \rho \propto r^{-0.47} $, while for model in the presence of thermal conduction, it follows 
$ \rho \propto r^{-1.7} $. This implied that when the thermal conduction is not considered in 
the flow, the profile of the density has a smoother trend at the inner region 

To show the existence of outflow based on a direct way, we use a trajectory approach 
in this work. We set 34 trajectory test particles staring from $ r = 0.2 r_{\rm _B} $ and our results 
clearly show the real wind trajectories. In principle, there exist particles that extend from starting 
points to the large radii above the Bondi radius and never cross $ r = 0.2 r_{\rm _B} $ twice.
After obtaining the trajectories, we also calculate the mass fluxes of inflow, outflow, as well as
real wind. In addition, to quantitatively investigate the wind, we calculated the ratio of mass flux of winds to 
the total outflow rate at $ r = 0.2 r_{\rm _B} $.  Our results show that about $ 80 \% $ of the total 
mass flux of outflow is real wind at this radius. Our quantitative calculation based on the 
mass fluxes and trajectory method can clearly confirm that the real wind exist in hot accretion flow 
and can reaches to large radii. Although in previous study of \citealt{Inayoshi et al. 2018} they found 
no outflow, here the key difference between our work and their work is we found 
that a very effective wind can be produced inside the Bondi radius. Hence, we conclude that the inward decrease 
of the mass accretion rate will be due to the wind rather than the convection at such large radii.

\citealt{Bu et al. 2016a, Bu et al. 2016b} also studied the accretion flow around Bondi radius. 
They find that after including the gravity of nuclear stars, the physics can be totally changed. 
Specifically, they find that in the presence of nuclear star gravity, the winds can not be produced locally. 
They also did some test and find that when the neclear star gravity is excluded, the winds can 
be generated locally around Bondi radius. Therefore, the result that the presence of winds near Bondi 
radius found in the present paper is not conflict with that of \citealt{Bu et al. 2016a, Bu et al. 2016b}.

Based on the above argument, for the hot accretion flow around the Bondi radius in a real galaxy, 
whether winds can be produced locally around Bondi radius depends on the presence or strength 
of  nuclear star gravity.  Around the Bondi radius, If the black hole gravity is significantly dominate 
the nuclear star gravity, we will expect strong winds generated around Bondi radius. However, if 
the nuclear star gravity dominates, no winds can be expected around Bondi radius.

There are several caveats in our study that will be improved in our future simulations. The first one is 
that we only consider the HD equations of hot accretion flow for our simulations. In a real accretion 
flow, angular momentum is transferred by Maxwell stress associated with MHD turbulence driven 
by MRI. Moreover, the magnetic field is one of the driving mechanisms for wind production. With 
this aim, we will preform the MHD simulations of the accretion flow at large radii. In our future  
work we will mainly focus on different magnetic field configurations on the structure of the hot 
accretion flow and compare the results with the HD simulations. A further simplification here is 
that we only considered the gravity of the central black hole. Our estimation
shows that in this range of radii the nuclear stars gravity might be taking into account.
Moreover, one-temperature fluid equations are considered here. In terms of hot accretion model, 
the ions are expected to be much hotter than the electrons (see \citealt{Rees et al. 1982}; 
\citealt{Yuan and Narayan 2014}). Thus, two different energy equations for electrons and ions should be solved.

\begin{acknowledgments}
We thank the anonymous referees for their thoughtful and constructive comments on an early version of the paper.
A.M. is supported by the National Natural Science Foundation of China (Grant No. 12150410308), 
foreign experts project (Grant No. QN2022170006L) and also the China Postdoctoral Science 
Foundation (grant No. 2020M673371). D.-F. B. is supported by the Natural Science Foundation 
of China (grant No. 12173065). M\v{C} is supported by the Polish NCN grant 2019/33/B/ST9/01564. 
M\v{C} was also supported by the ESF projects No. $\mathrm{CZ.02.2.69/0.0/0.0/18\_054/0014696}$.
F.Z.Z. is supported by the National Natural Science Foundation of China (grant No. 12003021), 
foreign experts project (Grant No. QN2022170005L) and also the China Postdoctoral Science 
Foundation (grant No. 2019M663664). L.M. is supported by the National Natural Science Foundation 
of China (grant No. 12171385). A.M. also acknowledges the support of Dr. X. D. Zhang at the 
Network Information Center of Xi’an Jiaotong University. The computation has made use of the 
High Performance Computing (HPC) platform of Xi’an Jiaotong University.
\end{acknowledgments}

\bibliography{sample631}{}
\bibliographystyle{aasjournal}



\end{document}